\documentclass[aps,showpacs,twocolumn]{revtex4}
\usepackage{graphics}
\begin{document}
\newcommand{\bstfile}{osa} 
\draft
\title{Experimental verification of a self-consistent theory of the first-, second-, and third-order (non)linear optical response}

\author{Mark G. Kuzyk} \email{kuz@wsu.edu} \affiliation{Department
of Physics and Astronomy; and Materials Science Program, Washington State University, Pullman,
Washington 99164-2814}

\author{Javier P\'erez-Moreno}
\email{Javier.PerezMoreno@fys.kuleuven.be} \affiliation{Department
of Chemistry, University of Leuven, Celestijnenlaan 200D, B-3001
Leuven, Belgium}

\author{Juefei Zhou}
\email{juefei.zhou@case.edu}
\affiliation{Department of Physics and Astronomy, Washington State University, Pullman,
Washington  99164-2814\footnote{{\em Current Address:} Department of Physics, Case Western Reserve University, 13900 Euclid Ave, Cleveland, OH 44106, USA}}

\author{Shiva K. Ramini} \email{rshiva@wsu.edu} \affiliation{Department
of Physics and Astronomy, Washington State University, Pullman,
Washington 99164-2814}

\author{Sheng-Ting Hung}
\email{ShengTing.Hung@fys.kuleuven.be} \affiliation{Department of
Chemistry, University of Leuven, Celestijnenlaan 200D, B-3001
Leuven, Belgium,}\affiliation{Department of Physics and Astronomy, Washington State
University, Pullman, Washington 99164-2814}

\author{Koen Clays}
\email{Koen.Clays@fys.kuleuven.be} \affiliation{Department of
Chemistry, University of Leuven, Celestijnenlaan 200D, B-3001
Leuven, Belgium,}\affiliation{Department of Physics and Astronomy, Washington State
University, Pullman, Washington 99164-2814}

\date{\today}

\begin{abstract}
We show that a combination of linear absorption spectroscopy, hyper-Rayleigh scattering, and a theoretical analysis using sum rules to reduce the size of the parameter space leads to a prediction of the two-photon absorption cross-section of the dye AF455 that agrees with two-photon absorption spectroscopy.  Our procedure, which demands self-consistency between several measurement techniques and does not use adjustable parameters, provides a means for determining transition moments between the dominant excited states based strictly on experimental characterization.  This is made possible by our new approach that uses sum rules and molecular symmetry to rigorously reduce the number of required physical quantities.
\end{abstract}

\pacs{42.65.An, 33.15.Kr, 11.55.Hx, 32.70.Cs}

\maketitle

\vspace{1em}

\section{Introduction}

There is a long history of using nonlinear-optical techniques to build an understanding of the mechanisms of light-matter interactions.  Given the availability of mostly single-wavelength lasers, early measurements used time domain studies to deconvolute mechanisms such as molecular reorientation and the electronic response in liquids\cite{hellw71.02,green82.12,green84.09,etche87.01} which were used to make fast optical gates,\cite{etche83.01} and in solids, for example, to study excitons in quasi-one-dimensional polymeric crystals.\cite{green87.01,green88.01}

One of the first attempts to use dispersion in the Optical Kerr Effect (OKE) to understand the nonlinear-optical response in liquids was based on a qualitative comparison of experimental results with the sum-over-states (SOS) expression for the nonlinear-optical susceptibilities\cite{kuzyk89.01} as calculated by Orr and Ward.\cite{orr71.01} More recently, OKE dispersion measurements and more sophisticated multi-state models have been applied to determining the transition moments between excited states in silicon phthalocyanine-monomethacrylate.\cite{vigil01.01}

The difficulty with such approaches is that they require either non-realistically simple models with only a couple of parameters; or, complex models that involve transition moments between excited states that can not be independently verified by experiment.  Furthermore, a set of parameters that successfully models one particular measurement is not often consistent with other independent measurements.

The sum rules are quantum mechanical identities that relate the dipole matrix elements and energies to each other; so, the SOS hyperpolarizability can be expressed in terms of a subset of the dipole matrix.\cite{kuzyk00.01,kuzyk01.01,kuzyk03.02} Indeed, the dipole-free SOS expression is derived by using the sum rules to eliminate all terms with dipole moment differences. Dipole-free expressions for both the hyperpolarizability, $\beta$,\cite{kuzyk05.02} and the second hyperpolarizability, $\gamma$,\cite{perez01.08} have been derived and shown to be mathematically equivalent to the standard SOS results.  These dipole-free expressions are key to significantly reducing the number of parameters required to model the nonlinear response.

In the present work, we use linear absorption spectroscopy to determine the energies and transition moments from the ground state of the octupolar chromophore commonly referred to as AF-455, and use hyper-Rayleigh scattering, the $D_{3h}$ symmetry of the molecule, and the dipole-free SOS expression for the hyperpolarizability to determine the transition moment between the two dominant states.  Using {\em no adjustable parameters}, we predict a two-photon absorption spectrum that agrees with experimental results.  This suggests that our approach may be a simple alternative that is straightforward to apply and yields self-consistent results that spans linear absorption, hyper-Rayleigh scattering, and two-photon absorption spectroscopy.

What makes our work unique is that all of the quantities required to predict the linear and nonlinear-optical response are determined experimentally, which is made possible using sum rules to reduce the number of required parameters.  Other approaches have been introduced that provide estimates of the nonlinear response from simple measurements.  For example, P\'erez-Moreno and coworkers have introduced a rule of thumb that provides a rough estimate of the resonant two-photon absorption (TPA) cross-section of a large set of molecules simply using the number of $\pi$ electrons.\cite{perez05.01} Rebane and coworkers used a theoretical approach based on the density matrix to show that parameters determined from a measurement of the linear absorption spectrum of a dipole transition can be used to determine the TPA cross-section at the one-photon absorption maximum.\cite{reban08.01}  For a broad range of molecules, the approach was shown to be yield TPA cross-sections that deviated at most by 50\% compared with measurements.  However, while these approximate techniques are useful    for estimating the TPA cross-section, they do not predict the dispersion of the TPA cross-section, nor do they predict the first hyperpolarizability.  In contrast, the approach presented here leads to an accurate prediction of the TPA spectrum.

\section{Theory}

Our approach begins by simplifying the analysis of the dispersion of the nonlinear-optical susceptibilities by providing a model that depends only on a reduced set of measurable molecular parameters.  In contrast, most studies reported in the literature rely on calculating and/or using as fit parameters transition energies between excited states, excited state dipole moments, and energies.

Because excited-state parameters can not be experimentally verified, and often many sets of parameters can yield reasonably good fits of the data to the theory, the conclusions based on such studies may not be sound.  Furthermore, one set of parameters will often provide a good fit to one experiment, but not to others.  To compound such problems, semi-empirical calculations of a nonlinear susceptibility measurement off-resonance does not correctly predict the resonant behavior.

We begin by introducing the SOS dipole-free expressions of the nonlinear-optical susceptibilities. These expressions reduce the number of parameters needed to describe the nonlinear-optical response by eliminating the dipole terms.  Then, we use the symmetries of an octupolar molecule to show that the two-fold degenerate first excited state can be expressed in a basis where the transition moment between these two states vanishes.  However, this basis demands that the dipole moments be non-vanishing.  This is of no consequence since we use the dipole-free expressions to calculate the nonlinear susceptibilities.  Finally, applying the sum rules to the dipole basis, we show that the four-level system can be expressed as a three-state model.

\subsection{Sum Rules and the Dipole-Free SOS Expressions}

The polarizability of a molecule along $\hat{x}$ for an incident field of frequency $\omega$ polarized along $\hat{x}$ is given by,\cite{orr71.01}
\begin{equation}\label{alpha}
\alpha_{xx}(\omega) = e^2 {\sum_n^{\infty}} ' \left[ \frac {x_{0n} x_{n0}} {E_{n0} - \hbar \omega} +  \frac {x_{0n} x_{n0}} {E_{n0} + \hbar \omega} \right],
\end{equation}
where $x_{nm}$ is the transition moment between states $n$ and $m$,\footnote{A complex molecule is made of many electrons, so the dipole moment operator is proportional to a sum over the position operators of all the electrons.  In matrix form, this can be expressed as
$$
\mu_{ij} = -e \sum_n^N \left< \psi_i \left| \vec{r}_n \right| \psi_j \right> \equiv -e \vec{r}_{ij}.
$$
In this paper, we will loosely refer to $\vec{r}$ as the dipole moment.} $E_{n0} = E_n -E_0$ is the energy difference between states, and where the prime denotes the fact that the ground state is excluded from the sum.  The energies are complex to account for damping.  Note that while the susceptibilities are tensors, for the sake of simplicity, we express only the largest tensor component.

The sum rules are calculated from the Schr\"odinger Equation using the closure identity.  For charges under the influence of electric and magnetic fields, the sum rules are exactly obeyed, and are derived from the matrix elements of commutators of the Hamiltonian $H$ and position $x$, which yields\cite{kuzyk01.01}
\begin{eqnarray}\label{therule}
& & \frac {2m} {\hbar^2} \sum_{n}  x_{ln} x_{np} \left(  E_n - \frac {1} {2} \left( E_p + E_l \right) \right) = N \delta_{l,p}. \nonumber \\
\end{eqnarray}
Note that the set of relationships given by Equation \ref{therule} with $l=p$ are commonly called the {\em sum rules}.  The more general form above with $l \neq p$  was developed to account for the off-diagonal components.\cite{kuzyk00.01,kuzyk01.01}  Equation \ref{therule} corresponds to an infinite set of equations, which are each labeled by distinct pairs of integers $l,p$.  The sum rules clearly show how the matrix elements of the position operator (proportional to the dipole matrix) and energy levels are intimately related to each other and therefore can not be independently adjusted.

The infinite set of equations embedded in the sum rules can be used to simplify the SOS expressions for the nonlinear-optical susceptibilities.  The off-diagonal sum rules can be used to express the ground and excited state dipole moments in terms of transition moments.\cite{kuzyk05.02}  As such, the dipole terms can be eliminated.\cite{kuzyk05.02} Using these relationships, the common SOS expression can be transformed into the dipole-free SOS expression, which for the largest diagonal component of the hyperpolarizability yields,\cite{kuzyk05.02}
\begin{eqnarray}\label{beta contracted}
\beta_{xxx} (\omega_1, \omega_2) & = & - e^3
P_{\omega_{1}, \omega_{2}} \\ \nonumber & \times &
{\sum_{m}^{\infty}} ' {\sum_{n  \neq m}^{\infty}} ' \frac {x_{0m}
x_{mn} x_{n0}} {D_{nm}^{-1} (\omega_1 , \omega_2)} \\ \nonumber & \times & \left[ 1 - \frac {D_{nm}^{-1} (\omega_1 , \omega_2)}
{D_{nn}^{-1} (\omega_1 , \omega_2)} \left( 2 \frac {E_{m0}} {E_{n0}}
- 1 \right) \right] ,
\end{eqnarray}
where,
\begin{eqnarray}\label{beta denominators}
&& P_{\omega_{1}, \omega_{2}} [D_{nm} (\omega_1 , \omega_2)]  = \nonumber \\ && \frac {1} {2 \hbar^2} \left[ \frac {1} { \left(\omega_{n0} - \omega_1 - \omega_2 \right) \left(\omega_{m0} - \omega_1 \right)} \right. \nonumber \\
&+&  \frac {1} {\left(\omega_{n0}^* + \omega_2 \right) \left(\omega_{m0} - \omega_1 \right)}  \nonumber \\
&+&  \frac {1} {\left(\omega_{n0}^* + \omega_2 \right) \left(\omega_{m0}^* + \omega_1 + \omega_2 \right)} \nonumber \\
&+& \left. \omega_1 \leftrightarrow \omega_2 \hspace{1em} \mbox{for the three previous terms} \right] ,
\end{eqnarray}
and where $\omega_{m0} = \omega_{m0}^0 - i \gamma_{m0}$.  $\hbar
\omega_{m0}^0$ is the energy difference between excited state $\left| m \right>$ and the
ground state and $\gamma_{m0}$ is half the natural linewidth for a
transition from state $m$ to the ground state. This expression is sometimes called the reduced SOS expression.

The second hyperpolarizability, as well as higher-order hyperpolarizabilities, can be transformed into a dipole-free form.  The resulting algebraic expressions are too complex to present here, but their form can be found in the literature.\cite{perez01.08}

\subsection{The Dipole Basis}

\begin{figure}
\includegraphics{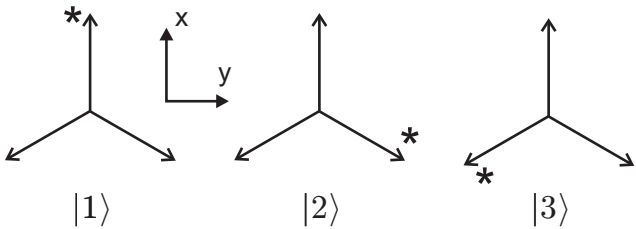}
\caption{The three degenerate states of an octupole with no charge hopping between sites.}
\label{fig:octupole}
\end{figure}

We begin by using the symmetries of an octupolar molecule with $D_{3h}$ symmetry to define the energy level diagram. It is typical to model an octupole with a two-fold degenerate excited state.\cite{Joffr92.01} This degeneracy can be understood by considering a basis in which the charge is placed on one of the three branches of the octupole as shown schematically in Figure \ref{fig:octupole}.  With no charge hopping, the states $\left| 1 \right>$, $\left| 2 \right>$, and $\left| 3 \right>$ are the three eigenstates of the system.

Using the approach described by Feynman,\cite{feynm65.03} we define a $120^o$ rotation by the operator $R$.  Because a rotation of the system by $360^o$ leaves the system invariant, $R^3 = 1$, and the eigenvalues of $R$ are clearly $1$, $\exp \left( 2 \pi i/3 \right)$, and $\exp \left( 4 \pi i/3 \right)$.  With hopping between nearest neighbors, the Hamiltonian is,
\begin{equation} \label{hopping hamiltonian}
H = {\bf 1 } \mathcal{E}_0 - \Delta R - \Delta R^{-1},
\end{equation}
where $\mathcal{E}_0$ is the energy without hopping and $ \Delta$ the hopping energy.  Since $H$ and $R$ commute, they share the same eigenfunctions, which are of the form:
\begin{eqnarray} \label{no dipole eigenfunctions}
\left| \psi_0 \right> & = & \frac {1} {\sqrt{3}} \left( \left| 1 \right> + \left| 2 \right> + \left| 3 \right>\right),\nonumber \\
\left| \psi_1 \right> & = & \frac {1} {\sqrt{3}} \left( \left| 1 \right> + e^{2 \pi i/3}\left| 2 \right> + e^{4 \pi i/3} \left| 3 \right>\right),\nonumber \\
\mbox{and} && \nonumber \\
\left| \psi_2 \right> & = & \frac {1} {\sqrt{3}} \left( \left| 1 \right> + e^{4 \pi i/3}\left| 2 \right> + e^{2 \pi i/3} \left| 3 \right>\right).
\end{eqnarray}
The corresponding Eigenenergies are,
\begin{eqnarray} \label{no dipole energies}
E_{0} & = & \mathcal{E}_{0} - 2 \Delta \nonumber \\
\mbox{and} && \nonumber \\
E_{1} & = & E_{2} = \mathcal{E}_0 + \Delta.
\end{eqnarray}

These three states describe the ground state and the two-fold degenerate excited state of the octupole.  If the excitation on branch $n$ designated by the state vector $\left| n \right>$ represents a dipole of magnitude $|\mu| = e \cdot l$, where $l$ is the distance between the center of the molecule and the end of one of the branches, then the dipole moment of state $i$ is given by,
\begin{equation} \label{no dipole moments}
\vec{r}_i = \left< \psi_i \right| \vec{r} \left| \psi_{i} \right> = \frac{l}{3} \left(\hat{1} + \hat{2} + \hat{3} \right) = 0,
\end{equation}
where the  unit vectors $\hat{1}$, $\hat{2}$, and $\hat{3}$ are along the three branches of the octupole represented by the state vectors $\left|1\right>$, $\left|2\right>$, and $\left|3\right>$, respectively, as diagrammed in Figure \ref{fig:octupole}; and, $- e \cdot l \cdot \hat{n}$ is the dipole moment of the branch represented by the state vector $\left| n \right>$.   Note that we have used
\begin{equation} \label{No overlap between branches}
\left< n \right| \vec{r} \left| m \right> = l \cdot \delta_{n,m} \cdot \hat{n}
\end{equation}
under the assumption that there is no overlap between the wavefunctions on different branches.  Because each state has no dipole moment, we call this the {\em octupolar basis}.

\begin{figure}
\includegraphics{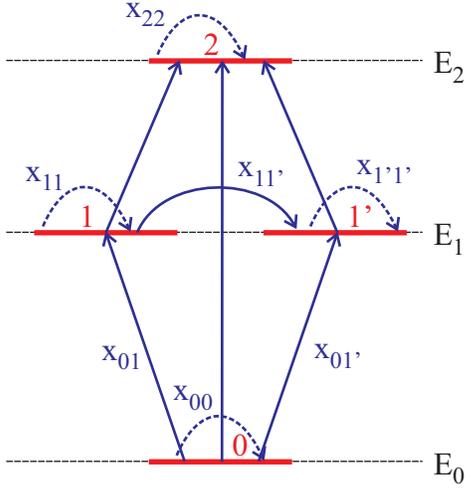}
\caption{The four-level model of an octupolar molecule.  All possible transition moments along the $\hat{x}$-direction are shown.}
\label{fig:4level}
\end{figure}

Since the three-level model may not be sufficient to describe an octupole, we add a third excited state.  The properties of this state are completely general and there are no restrictions on its dipole moment. Figure \ref{fig:4level} shows the energy level diagram of the four-state system that we will use for our analysis and labels all of the elements of the dipole matrix.  Note that the degenerate states are relabeled $1$ and $1'$ to stress that these states are degenerate with energy $E_1$.

In the octupolar basis, $\vec{r}_{00}$, $\vec{r}_{11}$, and $\vec{r}_{1'1'}$ vanish.  It is simple to show that $\vec{r}_{11'} \neq 0$, and will be of the form,
\begin{equation}\label{tansitions}
\vec{r}_{11'} \propto \left(\hat{1} + e^{2 \pi i/3} \hat{2} +  e^{4 \pi i/3}\hat{3} \right) \neq 0.
\end{equation}
Since any linear combination of two degenerate energy eigenvectors is an energy eigenvector, it is possible to chose linear combinations of $\left| \phi_1 \right> $ and $\left| \phi_1' \right> $ to form two orthogonal energy eigenvector with $\hat{x} \cdot \vec{r}_{11'}=0$; but then, $\vec{r}_{11}$ and $\vec{r}_{1'1'}$ will no longer vanish.  In this new basis, the vectors in the degenerate subspace are given by,
\begin{eqnarray} \label{dipole eigenfunctions}
\left| \phi_1 \right> & = & \frac{1}{\sqrt{2}} \left( \left| \psi_{1} \right> \right) + \left( \left| \psi_{2} \right> \right) = \frac {1} {\sqrt{6}} \left( 2 \left| 1 \right> - \left| 2 \right> - \left| 3 \right>\right).\nonumber \\
\left| \phi_{1'} \right> &=&\frac{1}{i \sqrt{6}} \left( \left| \psi_{1} \right> \right) - \left( \left| \psi_{2} \right> \right) = \frac {1} {\sqrt{2}} \left(\left| 2 \right> - \left| 3 \right>\right).
\end{eqnarray}
We call this the {\em dipole basis}.  Note that $\left| \phi_0 \right> = \left| \psi_0 \right> $ and $\left| \phi_2 \right> = \left| \psi_3 \right> $.

The dipole moments of state $\left| \phi_1 \right>$ and $\left| \phi_{1'} \right>$ in the dipole basis are then given by,
\begin{eqnarray} \label{dipole moments}
 \vec{r}_{11} & \propto & \frac {1} {6} \left( 2 \hat{1} + \hat{2} + \hat{3} \right) \propto \hat{x},\nonumber \\
\vec{r}_{1'1'} & \propto & \frac {1} {2} \left(+\hat{2} + \hat{3} \right) \propto - \hat{x},
\end{eqnarray}
where we have used the coordinate convention shown in Figure \ref{fig:octupole}. Clearly, the dipoles in the two degenerate states point in opposite directions.  It is straightforward to show that the transition moment $\vec{r}_{11'}$ has no $\hat{x}$-component in the dipole basis:
\begin{equation}
\vec{r}_{11'} \propto  \frac{1}{i 2\sqrt{3}} \left( -\hat{2} + \hat{3} \right) \propto \hat{y}.
\end{equation}

Finally, in the dipole basis the transition dipole moment $\vec{r}_{01'}$ has not $\hat{x}$-component:
\begin{equation}
\vec{r}_{01'} \propto  \frac{1}{\sqrt{6}} \left( \hat{2} - \hat{3} \right) \propto \hat{-y}.
\label{eq:x01prime}
\end{equation}

\subsection{Applying the sum rules to the dipole basis}

Our approach is to determine the hyperpolarizability and second hyperpolarizability along $\hat{x}$, that is, to calculate $\beta_{xxx}$ and $\gamma_{xxxx}$, and then to use octupolar symmetry to determine the other components.  As such, we need only consider the one-dimensional sum rules.  Furthermore, we will assume that four states are sufficient to describe the nonlinear response, and will therefore only consider four-state sum rules.

We evaluate the sum rules in the dipole basis, which is chosen deliberately to meet the condition $x_{11'} = 0$.  Equation \ref{eq:x01prime} yields $x_{01'}=0$.  To summarize, the dipole basis demands: (1) the {\em no coupling condition}, $x_{11'} = 0$; (2) the transition moment from the ground to primed state vanishes, $x_{01'}=0$; and (3) the ground state moment vanishes, $x_{00}=0$.  We note that the dipole basis does not impose any other restrictions on the moments.  However, as we show below, the sum rules impose an additional set of conditions.

The sum rule $l=0$ and $p=1'$ yields,
\begin{eqnarray}\label{l=0,p=1'}
&-& x_{00} x_{01'} E_{10} + x_{01} x_{11'} E_{10} + x_{01'} x_{1'1'} E_{10} \nonumber \\
&+& x_{02} x_{21'} \left(2 E_{20} - E_{10} \right) = 0,
\end{eqnarray}
which in the dipole basis (with $x_{00} = 0 $, $x_{01'}=0$ and $x_{11'}=0$) yields
\begin{equation}\label{l=0,p=1' dipole}
x_{02} x_{21'} \left(2 E_{20} - E_{10} \right) = 0.
\end{equation}
Since $(2 E_{20} - E_{10}) >0$, the implication is that $x_{02}$ or $x_{21'}$ must vanish.  We choose $x_{21'}=0$ because it is in agreement with the sum rule $l=1$ and $p=1'$, which yields
\begin{eqnarray}\label{l=1',p=1}
&-& x_{10} x_{01'} E_{10} + x_{11} x_{11'} E_{11} \nonumber \\
&+& x_{11'} x_{1'1'} E_{11} + x_{12} x_{21'} E_{21} = 0.
\end{eqnarray}
As a consequence of the fact that $E_{11}=0$, 
\begin{equation}\label{l=1',p=1 dipole}
x_{12} x_{21'} \left( E_{20} - E_{10} \right) = 0.
\end{equation}
Since $E_{20} - E_{10} > 0$, we conclude that $x_{21'}=0$, which is in agreement with Equation \ref{l=0,p=1' dipole}.

The net result is that all transitions that include state $1'$ are forbidden.  We call this {\em channel blocking}.  As a result of channel blocking, the four-level system can be described by a three-level model.  Note that we could have alternatively made the choice that state $1$ is the excluded state (i.e. $1 \Leftrightarrow 1'$).  This relabeling is equivalent to rotating the coordinate system by $90^o$.

We note that while the choice of one basis over another results in different dipole moments and energies of the degenerate states, each basis will yield the same observables - such as the absorption spectrum, hyperpolarizability, etc.  So, we choose the dipole basis because it is the most convenient: the number of parameters required to describe the nonlinear response is reduced and the dipole-free form of the nonlinear susceptibilities makes the values of the ground and excited state dipole moments irrelevant.

\subsection{The Polarizability}

In the four-state model, the sum rule with $l=0$ and $p=0$ yields,
\begin{equation}\label{l-0,p=0}
\left| x_{01}\right|^2 E_{10} + \left| x_{01'}\right|^2 E_{10} + \left| x_{02}\right|^2 E_{20} = \frac {\hbar^2 N} {2m}.
\end{equation}
Since the transition moment to the second excited state and the energies of the degenerate states and second excited state are independent of the basis,
\begin{equation}\label{l-0,p=0}
\left| x_{01}\right|^2 + \left| x_{01'}\right|^2 = \frac {\hbar^2 N} {2m E_{10}} - \left| x_{02}\right|^2 \frac {E_{20}} {E_{10}}= \mbox{constant}.
\end{equation}

The polarizability near resonance with the degenerate state ($E_{10} \approx \hbar \omega$) is give by Equation \ref{alpha},
\begin{equation}\label{alpha-res}
\alpha \approx e^2  \frac {\left| x_{01}\right|^2 + \left| x_{01'}\right|^2} {E_{10} - \hbar \omega} .
\end{equation}

The area under the peak in the absorption spectrum near $E_{10} = \hbar \omega$ is proportional to $\left| x_{01}\right|^2 + \left| x_{01'}\right|^2$, and no measurement can separate the individual contributions of each degenerate state.  Since the sum rules demand that any choice of basis in the degenerate subspace must yield $\left| x_{01}\right|^2 + \left| x_{01'}\right|^2 = \mbox{constant}$, all bases will lead to the same observed linear absorption spectrum.  Thus, the choice of basis has no effect on any observable.  The dipole basis is the particular choice where $x_{01'}=0$ so all of the oscillator strength in the subspace is concentrated in $x_{01}$; and, the area under the linear absorption peak centered at $E_{10}$ will be proportional to $\left| x_{01}\right|^2 E_{10}$.

\subsection{The Hyperpolarizability in the Dipole Basis}

In this section, we impose octupolar $D_{3h}$ symmetry and the sum rules to calculate the hyperpolarizability using the dipole-free expression in the dipolar basis.  In analogy to the rigorous proof that the polarizability is independent of basis, the same will hold for the hyperpolarizability, which we state here without proof.

\begin{figure}
\includegraphics{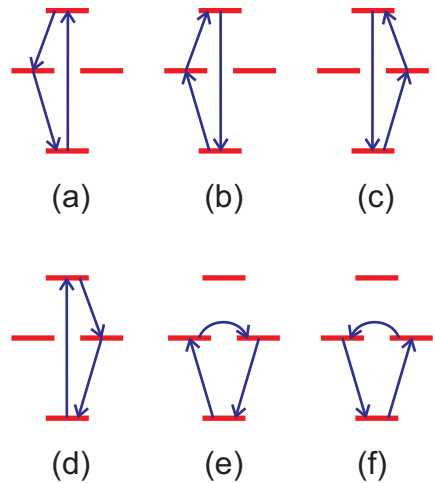}
\caption{All possible contributions to the dipole-free form of the hyperpolarizability in a four-state system.}
\label{fig:betaGraphs}
\end{figure}

Recall that the dipole-free SOS expression for the hyperpolarizability has numerators of the form $x_{0n} x_{nm} x_{m0}$ where $m\neq n$ and $m\neq 0$ for all $m$ and $n$.  It is convenient to represent all possible virtual transitions graphically, as shown in Figure \ref{fig:betaGraphs}

In the dipole basis, diagrams (e) and (f) are forbidden because of the {\em no coupling condition}.  Diagrams (c) and (d) are forbidden by {\em channel blocking}

The only non-vanishing terms represented by diagrams (a) and (b) represent the three-level dipole-free SOS expression for $\beta_{xxx}$.  We note that the numerators associated with these diagrams are complex conjugates of each other.  The contribution to the hyperpolarizability of one diagram is the same as the other one with states $1$ and $2$ interchanged.

\subsection{The Second Hyperpolarizability in the Dipole Basis}

\begin{figure}
\includegraphics{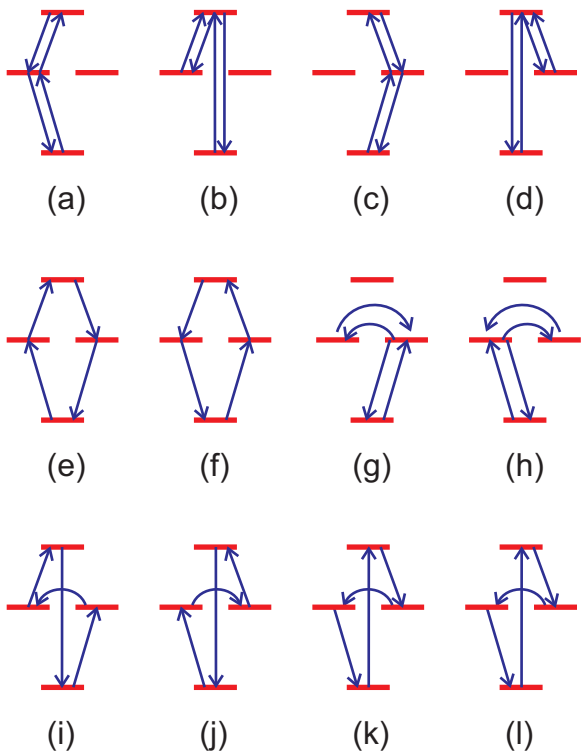}
\caption{All possible contributions to the dipole-free form of the second hyperpolarizability in a four-state system of the Type I terms.}
\label{fig:gammaGraphs}
\end{figure}

In this section, we impose octupolar $D_{3h}$ symmetry and the sum rules to calculate the second hyperpolarizability using the dipole-free expression in the dipolar basis.  The dipole-free SOS expression for the second hyperpolarizability has two types of terms.  In the first type, called Type I, the numerators are of the form $x_{0n} x_{nm} x_{ml} x_{l0}$ where $m\neq n$ and $m\neq l$.  Also, no index can represent the ground state, $0$.  It is convenient to represent all possible virtual transitions graphically, as shown in Figure \ref{fig:gammaGraphs}

Diagrams (c) through (f) are forbidden by channel blocking.  Diagrams (g) through (l) are forbidden because of the no-coupling condition.  As in the case of the hyperpolarizability, the non-vanishing diagrams (a) and (b) represent the two terms in the three-level dipole-free SOS expression for the second hyperpolarizability.

\begin{figure}
\includegraphics{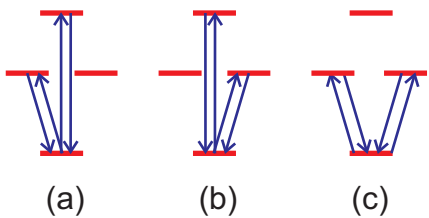}
\caption{All possible contributions to the dipole-free form of the second hyperpolarizability in a four-state system of the Type II terms.}
\label{fig:gammaGraphs2}
\end{figure}

Next we consider Type II terms, as shown in Figure \ref{fig:gammaGraphs2}.  Diagrams (b) and (c) are forbidden by channel blocking.  The only non-vanishing term is once again the three-level dipole-free Type II term.

\section{Results and Discussion}

The dipole basis, when used with the dipole-free SOS expression, leads to a model of the linear and nonlinear susceptibilities that depends only on the energies of the two excited states $E_{10}$, $E_{20}$; their widths; and the transition moments $x_{01}$, $x_{02}$, and $x_{12}$.  The quantities $E_{10}$, $E_{20}$, $x_{01}$, and $x_{02}$ can be determined using the linear absorption spectrum.  The only remaining quantity is $x_{12}$.  The transition moment between the two excited states can be determined from a measurement of $\beta$ at one wavelength provided that the other quantities are known {\em a priori}.  From these measured quantities, we will show that all-optical and nonlinear-optical properties are predicted with no adjustable parameters.

\begin{figure}
\includegraphics{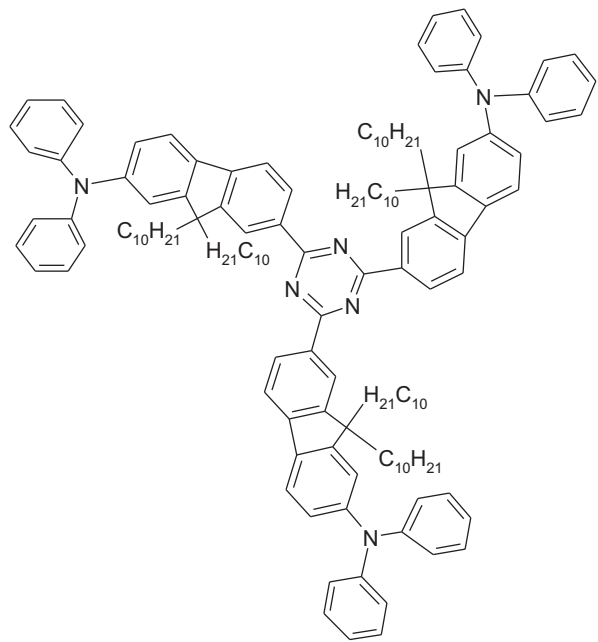}
\caption{The AF455 molecule.  Note its $D_{3h}$ symmetry.}
\label{fig:AF455}
\end{figure}
We apply this technique to the octupolar molecule AF-455 in solution, which as shown in Figure \ref {fig:AF455}, has $D_{3h}$ symmetry.  The absorption spectrum is used to determine its polarizability using the fact that with appropriate choice of molecular coordinate system, $\alpha_{xx} = \alpha_{yy}$ and $\alpha_{xy} = \alpha_{yx} = 0$.  Our analysis holds generally for any octupolar molecule with $D_{3h}$ symmetry, which we will refer to simply as an octupole.  An isotropic solution of octupoles will have an isotropic polarizability of $\left< \alpha \right> = 2 \alpha_{xx}/3$.  Thus, from an absorption spectrum, the polarizability along $\hat{x}$ can be directly determined.

\begin{figure}
\includegraphics{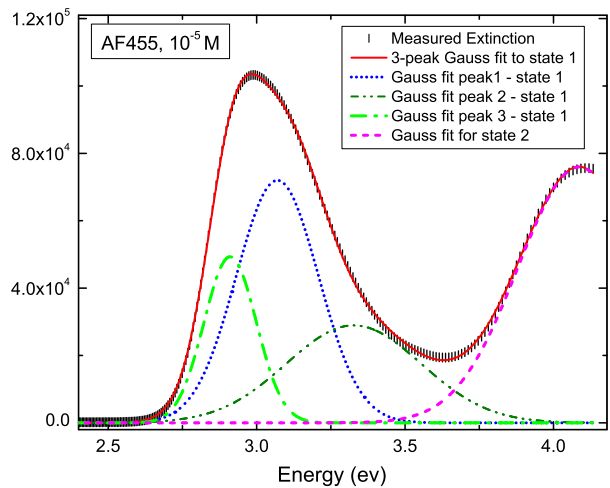}
\caption{Extinction spectrum of AF455 and fit to four Gaussians; three to fit the first excited state and one for the second excited state.}
\label{fig:extinction}
\end{figure}

The linear absorption spectrum of a sample is obtained by measuring the transmittance
of a broad spectrum source passing through a cuvette of 1cm path length, and containing a
solution of the sample in solvent.  The spectrum is referenced to the transmittance through pure
solvent in an identical cuvette.  Figure \ref{fig:extinction} shows the measured extinction spectrum of AF455 in tetrahydrofuran (THF).

The two peak positions are used to determine the energies $E_{10}$ and $E_{20}$; and , the width at half maximum for each peak is used to determine $\Gamma_{10}$ and $\Gamma_{20}$.  The absorption spectrum is fit to four Gaussian peaks; three for the first excited state and one for the second excited state as shown in Figure \ref{fig:extinction}.  The sum of the areas of the first three peaks is used to calculate the transition moment to the first excited state, using the method described by Tripathi et al,\cite{tripa04.01} while the area of the fourth Gaussian is used to get the transition moment to the second excited state.  The solid curve is the sum of all four peaks, and shows that the four-Gaussian theory provides a good fit to the data.  The results are summarized in Table \ref{tab:fitParameters}.
\begin{table*}
\begin{eqnarray*}
\begin{array}{c c c c c c c c c c }
\hline
\mbox{Fitting Method} & E_{10} \mbox{eV} & E_{20} \mbox{(eV)} & |\mu_{10}| \mbox{(D)} & |\mu_{20}| \mbox{(D)} & \Gamma_{10} \mbox{(eV)} & \Gamma_{20} \mbox{(eV)} & |\mu_{12}|^{\mbox{upper}} \mbox{(D)} & |\mu_{12}|^{\mbox{middle}} \mbox{(D)} & |\mu_{12}|^{\mbox{lower}} \mbox{(D)}   \\
\hline \hline
\mbox{\bf 3 Gaussian} & 3.0 & 4.1 & 12.6 & 9.4 & 0.22 & 0.35 & 15.5 & 12.6 & 9.7 \\ \hline
\end{array}
\end{eqnarray*}
\caption{Experimental parameters (from the linear absorption spectrum) and $|\mu_{12}|$ measured using HRS. The range of HRS values reflect experimental uncertainties.  All quantities are dressed values.\label{tab:fitParameters}
}
\label{tab:betas}
\end{table*}

The hyperpolarizability is determined through hyper-Rayleigh scattering (HRS), using a femtosecond pulsed laser (Millennia X + Tsunami with a lock-to-clock system that ensures 80Mhz pulsed output) at fundamental wavelength of $\lambda = 800nm$ in conjugation with a low frequency lock in amplifier and a signal generator. Details can be found elsewhere.\cite{clays91.01,clays92.01}  The measured HRS hyperpolarizability yields,
\begin{equation}\label{betaHRS}
\left< \beta_{HRS}^2 \right> = \frac {8} {21} \beta_{xxx}^2 .
\end{equation}
The octupolar symmetry of the compound (which is used in Equation \ref{betaHRS} to determine $\beta_{xxx}$ directly from an HRS measurement) was confirmed by the depolarization measurements.  Note that the other components of the $\beta$ tensor are given by,\cite{kieli71.01}
\begin{equation}\label{betaOctusymmetry}
\beta_{xxx} = - \beta_{xxy} = - \beta_{xyx} = - \beta_{yxx} ,
\end{equation}
while all other tensor components vanish. A demodulation technique is used to determine a fluorescence-free value to insure that only the hyperpolarizability is being measured.\cite{Olbre98.01}  We note that for AF455, no fluorescence contribution was found at 400nm, which leads to a more accurate determination of the first hyperpolarizability than is possible with the demodulation technique when fluorescence is present.

The dipole-free expression for the $xxx$-component of the first hyperpolarizability for a three-state model is given by\cite{kuzyk05.02}
\begin{eqnarray}\label{diople-free beta}
\beta_{DF} (\omega_1, \omega_2) & = & - e^{3}
{\sum_{m=1}^{2}} {\sum_{n  \neq m}^{2}} \frac {x_{0m}
x_{mn} x_{n0}} {D_{nm}^{-1} (\omega_1 , \omega_2)} \\ \nonumber
& \times & \left[ 1 - \frac {D_{nm}^{-1} (\omega_1 , \omega_2)}
{D_{nn}^{-1} (\omega_1 , \omega_2)} \left( 2 \frac {E_{m0}} {E_{n0}}
- 1 \right) \right],
\end{eqnarray}
where $D_{nm}^{-1} (\omega_1 , \omega_2)$ is a dispersion function that depends only on the energies of the first two excited states, the two photon frequencies, and the widths of the two states.  In the HRS measurement, $\omega_1 = \omega_2 = \omega$.  We note that for an infinite number of states, the standard SOS and dipole-free expressions are rigorously identical; but when truncated to three states, the two are different.  (Note that our sum rule/symmetry analysis represents a four-level system with a three-level model without loss of information.) It is not possible {\em a priori} to know which expression is more accurate.  Our strategy is to use the dipole-free expression because it does not require knowledge of the dipole moments.  The merits of this approach will be judged by the predictive capability of the model.

The linear absorption measurement determines all the transition moments, energies and widths except for $x_{12}$.  Thus, given that the first hyperpolarizability is measured at a known frequency $\omega$, Equation \ref{diople-free beta} can be inverted to solve for $\mu_{12} = -e x_{12}$.  Table \ref{tab:fitParameters} shows all of the parameters determined from linear absorption spectroscopy and one measurement of the first hyperpolarizability.  Note that the transition dipole moment is related to the position matrix elements, $\mu_{nm} \equiv -e x_{nm}$.  The three values of $\mu_{12}$ listed in Table \ref{tab:fitParameters} represent the uncertainty range due to uncertainty in the HRS measurement.

\begin{figure}
\includegraphics{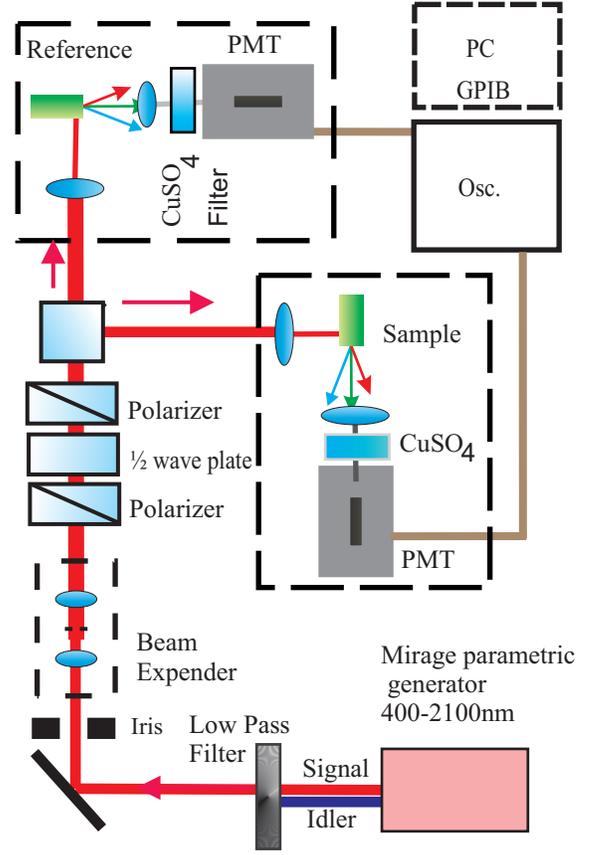}
\caption{The experiment used for measuring the two-photon absorption cross-section.}
\label{fig:TPASETUP}
\end{figure}

Figure \ref{fig:TPASETUP} shows the experiment used for determining the two-photon absorption cross-section using the measured two-photon fluorescence power.  This technique was developed by Xu and Web.\cite{xu96.02}  Details of how the data is related to the two photon cross-section can be found in the original paper.\cite{xu96.02}  The advantage of this experiment is that it is a reliable method for determining the TPA cross-section\cite{Oulia01.01} and is not as susceptible to excited state absorption as is nonlinear transmission.\cite{suthe05.01,belfi07.01}  Here we briefly describe those issues that are particular to our implementation of the technique.

The sample solution is prepared by adding 0.0958 gram of AF455 crystals as received
from Wright Patterson Air Force Base to 200ml optical spectrum grade THF, in a clean flask at room temperature.  The mixture is agitated in an ultrasonic water bath for thirty minutes to make a uniform solution. Two quartz cuvettes (ordered as a matched pair) are filled, and labeled S1 and S2, respectively, are filled to 4/5 full with
the uniform solution. These two identical samples are used to calibrate the collection
efficiency of the reference and sample arms of the TPA experiment.

The reference solution is made from 200ml of 100M Rhodamine B solution by adding
0.00958 grams of Rhodamine B powder to 200ml optical spectrum grade methanol followed
by agitation in an ultrasound bath for 30 minutes at room temperature.  A clean cuvette
is filled with this solution to 4/5 full and capped tightly.

A sample is excited with a laser beam, and the two-photon fluorescence signal (integrated over all fluorescence wavelengths) is detected with an RCA C31034A-02 photomultiplier tube (PMT).  This PMT is of high quantum efficiency over a broad range of wavelengths throughout the visible.  Filters are used to remove any wavelengths corresponding to the pump energy or less, leaving only the fluorescence spectrum due to two-photon absorption.  The time-integrated current from the PMT is proportional to the number of two-photon absorptions.  A typical run uses the average over many lasers pulses of the 10 Hz laser source to decrease noise.  The reference is used to take into account laser fluctuations and is used as a standard for determining the absolute two-photon absorption cross-section.

For molecules of $D_{3h}$ octupolar symmetry, the tensor components are related to each other according to,\cite{kieli71.01}
\begin{equation}\label{gammasymmetry}
\gamma_{xxxx} = \gamma_{yyyy} = 3 \gamma_{xxyy} = 3 \gamma_{xyxy} = \mbox{ etc}\dots
\end{equation}
Thus, knowledge of the value of $\gamma_{xxxx}$ allows one to determine all other components using Equation \ref{gammasymmetry}. We use the {\em dipole-free expression} to calculate $\gamma_{xxxx}$ because it does not require knowledge of the ground and excited state dipole moments.\cite{perez01.08}  We do not present the dipole-free-expression here, but rather refer the reader to the literature.\cite{perez01.08}

\begin{figure}
\includegraphics{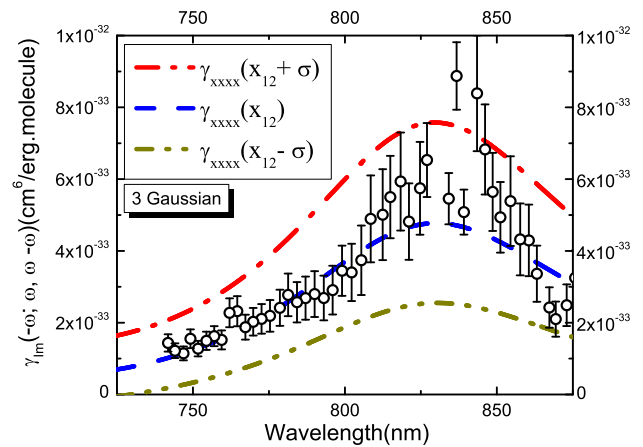}
\caption{The measured two-photon absorption spectrum (points) and the theoretical curve (middle), where all quantities are converted into the imaginary part of $\gamma_{xxxx}$.  The upper and lower curves represent the uncertainty introduced due to the uncertainty in the HRS-determined transition moment $\mu_{12}$.  Note that there are {\bf \em no adjustable parameters} in the theory.}
\label{fig:TPA_3G_mod_vs_data}
\end{figure}

Figure \ref{fig:TPA_3G_mod_vs_data} shows the theoretically-calculated value of the imaginary part of $\gamma_{xxxx}$ as a function of wavelength using the measured values of the dipole moment matrix elements, energies, and widths shown in Table \ref{tab:fitParameters}.  The upper and lower curves show the uncertainty range due to the experiential uncertainty in determining the excited state transition moment $\mu_{12}$ from HRS.

The two-photon absorption spectrum, as measured with the two-photon fluorescence experiment, yields an isotropic average over all tensor components of the two-photon absorption spectrum.  Given the $D_{3h}$ symmetry of the molecule, the isotropic value can be related to the individual tensor components.  Thus, the measured TPA cross-section can be related directly to the imaginary part of  $\gamma_{xxxx}$.  The measured values are shown as points with error bars in Figure \ref{fig:TPA_3G_mod_vs_data}.  We note that since the samples used in all measurements are liquid solutions, all quantities such as the polarizability, hyperpolarizability, transition dipole moments, etc. are dressed values.\cite{kuzyk98.01}  Vacuum quantities can be determined using the appropriate local field models.\cite{kuzyk98.01}

The theoretical spectrum and the data are in good agreement and all but one of the data points fall within the error band of the predicted spectrum.  Thus, the approach of using the dipole-free expressions for the first and second hyperpolarizability in the dipole basis provides a theoretical description that is consistent with three sets of measurements.  In particular, the predicted TPA spectrum through the two-photon resonance is consistent with the data.  This suggests that our approach may be useful in modeling the dispersion of the linear and nonlinear susceptibilities of octupolar molecules with $D_{3h}$ symmetry.

In light of the fact AF455 is a complex molecule, it may appear somewhat surprising that the predicted dispersion of the TPA cross-section using a four-level model is in such good agreement with the data.  This may be due to two factors.  First, the HRS measurement was determined near two-photon resonance, where the TPA peak is measured so the contributions of the first excited states are heavily weighted and dominate the response.  In contrast, an off-resonant HRS measurement potentially includes contributions from the tails of many higher-energy excited states, thus yielding an inaccurate determination of the transition moment $\mu_{12}$.  On-resonance measurement insures that the influence of the transition moment $\mu_{12}$ is large.  Secondly, since the TPA spectrum is measured only near the two-photon resonance, the same set of states are being probed.  Finally, it is possible that the observed agreement is a coincidence.  Similar studies of other octupolar systems would determine the general applicability of our method.

It is instructive to apply the same approach to the three-level model using the standard SOS expressions under the assumption that the three states are non-degenerate and have no dipole moment.  Using the value of $x_{12}$, determined form HRS measurements and the standard SOS expression form the hyperpolarizability, the theoretical value of the second hyperpolarizability predicts a two-photon absorption spectrum that is two-orders of magnitude larger than the measured one.  This illustrates how the typical approach can lead to inaccurate results, and supports the validity of our approach.

\section{Conclusion}

We have introduced an approach that combines measurements, symmetries, and sum rules to fully characterize the important states of a molecule that allows all of the linear and nonlinear susceptibilities to be accurately modeled with no adjustable parameters.  Our approach is general in that it can be applied to any molecule of any symmetry class.  In the present work, we have illustrated this approach for an octupolar molecule of $D_{3h}$ symmetry that is modeled using three excited states.  In addition to the sum rules, we have used the dipole basis and the dipole-free forms of the SOS expressions for the first and second hyperpolarizabilities, which do not require knowledge of the ground and excited state dipole moments.

We have found that the parameters in the four-level SOS model for a system with $D_{3h}$ symmetry can be fully determined using linear absorption spectroscopy and one near-resonant HRS measurement.  A key to reducing the number of measurements required is the use of symmetries and sum rules.  This approach bridges the gulf between the two-level model, which misses important states, and multilevel models, that use adjustable parameters to fit the data or results of semi-empirical calculations that can not be directly validated by experiment.\cite{fu07.01,fu07.02}  More importantly, our approach of combing theory with experiment leads to a small set of measured quantities that accurately predict the polarizability, hyperpolarizability, and second hyperpolarizability.

Our approach is generalizable to other systems using an analogous approach, provided that symmetries of the molecule exist that allow tensor components of the first (and second) hyperpolarizabilities to be related to each other.  The sum rules, used in conjunction with dipole-free SOS expressions, can then be used to simplify the model so that a reduced number of quantities are required to fully characterize a molecule.

\section{Acknowledgements}

MGK thanks the National Science Foundation (ECS-0756936) and Wright Patterson Air Force Base for generously supporting this work. We thank Wright Patterson AFB for supplying the AF455 chromophore used in these studies.  JPM acknowledges the Fund for Scientific Research Flanders (FWO).  KC acknowledges FWO grant G.0312.08

\bibliographystyle{\bstfile}

\begin{thebibliography}{10}
\newcommand{\enquote}[1]{``#1''}
\expandafter\ifx\csname url\endcsname\relax
  \def\url#1{{#1}}\fi
\expandafter\ifx\csname urlprefix\endcsname\relax\def\urlprefix{}\fi

\bibitem{hellw71.02}
R.~W. Hellwarth, A.~Owyoung, and N.~George, \enquote{{Origin of the Nonlinear
  Refractive Index of Liquid $CCl_4$},} Phys. Rev. A {\bf 4}, 2342--2347
  (1971).

\bibitem{green82.12}
B.~I. Greene and R.~C. Farrow, \enquote{{Direct Measurement of a Subpicosecond
  Birefringent Repsonse in $CS_22$},} J. Chem. Phys. {\bf 77}, 4779--4783
  (1982).

\bibitem{green84.09}
B.~I. Greene, P.~A. Fleury, J.~H.~L.~Carter, and R.~C. Farrow,
  \enquote{{Microscopic Dynamics in Simple Liquids by Subpicosecond
  Birefringences},} Phys. Rev. A {\bf 29}, 271--4 (1984).

\bibitem{etche87.01}
J.~Etchepare, G.~Grillon, G.~Hamoniaux, A.~Antonetti, and A.~Orszag,
  \enquote{{Molecular Dynamics of Liquid Benzene via Femtosecond Pulse Laser
  Excitation},} Rev. Phys. Appl. {\bf 22}, 1749--53 (1987).

\bibitem{etche83.01}
J.~Etchepare, G.~Grillon, A.~Migus, J.~L. Martin, and G.~Hamoniaux,
  \enquote{{Efficient Femtosecond Optical Kerr Shutter},} Appl. Phys. Lett.
  {\bf 43}, 406--407 (1983).

\bibitem{green87.01}
B.~I. Greene, J.~Orenstein, R.~R. Millard, and L.~R. Williams,
  \enquote{{Nonlinear Optical Response of Excitons Confined to One Dimension},}
  Phys. Rev. Lett. {\bf 58}, 2750--53 (1987).

\bibitem{green88.01}
B.~I. Greene, J.~F. Mueller, J.~Orenstein, D.~H. Rapkine, S.~Schmitt-Rink, and
  M.~Thakur, \enquote{{Phonon-Mediated Optical Nonlinearity in
  Polydiacetylene},} Phys. Rev. Lett. {\bf 61}, 325--28 (1988).

\bibitem{kuzyk89.01}
M.~G. Kuzyk, R.~A. Norwood, J.~W. Wu, and A.~F. Garito, \enquote{{Frequency
  dependence of the optical Kerr effect and third-order electronic
  nonlinear-optical processes of organic liquids},} J. Opt. Soc. Am. B {\bf 6},
  154--64 (1989).

\bibitem{orr71.01}
B.~J. Orr and J.~F. Ward, \enquote{{Perturbation Theory of the Non-Linear
  Optical Polarization of an Isolated System},} Molecular Physics {\bf 20},
  513--526 (1971).

\bibitem{vigil01.01}
S.~R. Vigil and M.~G. Kuzyk, \enquote{{Absolute molecular optical Kerr effect
  spectroscopy of dilute organic solutions and neat organic liquids},} J. Opt.
  Soc. Am. B {\bf 18}, 679--691 (2001).

\bibitem{kuzyk00.01}
M.~G. Kuzyk, \enquote{{Physical Limits on Electronic Nonlinear Molecular
  Susceptibilities},} Phys. Rev. Lett. {\bf 85}, 1218 (2000).

\bibitem{kuzyk01.01}
M.~G. Kuzyk, \enquote{{Quantum limits of the hyper-Rayleigh scattering
  susceptibility},} IEEE Journal on Selected Topics in Quantum Electronics {\bf
  7}, 774 --780 (2001).

\bibitem{kuzyk03.02}
M.~G. Kuzyk, \enquote{{Erratum: Physical Limits on Electronic Nonlinear
  Molecular Susceptibilities},} Phys. Rev. Lett. {\bf 90}, 039\,902 (2003).

\bibitem{kuzyk05.02}
M.~G. Kuzyk, \enquote{{Compact sum-over-states expression without dipolar terms
  for calculating nonlinear susceptibilities},} Phys. Rev. A {\bf 72}, 053\,819
  (2005).

\bibitem{perez01.08}
J.~P\'{e}rez-Moreno, K.~Clays, and M.~G. Kuzyk, \enquote{{A new dipole-free
  sum-over-states expression for the second hyperpolarizability},} J. Chem.
  Phys. {\bf 128}, 084\,109 (2008).

\bibitem{perez05.01}
J.~P\'{e}rez~Moreno and M.~G. Kuzyk, \enquote{{Fundamental limits of the
  dispersion of the two-photon absorption cross section},} J. Chem. Phys. {\bf
  123}, 194\,101 (2005).

\bibitem{reban08.01}
A.~Rebane, N.~S. Makarov, M.~Drobizhev, B.~Spangler, E.~S. Tarter, B.~D.
  Reeves, C.~W. Spangler, F.~Meng, and Z.~Suo, \enquote{{Quantitative
  Prediction of Two-Photon Absorption Cross Section Based on Linear
  Spectroscopic Properties},} J. Phys. Chem. C {\bf 112}, 7997--8004 (2008).

\bibitem{Joffr92.01}
M.~Joffre, D.~Yaron, J.~Silbey, and J.~Zyss, \enquote{{Second Order Optical
  Nonlinearity in Octupolar Aromatic Systems},} J . Chem. Phys. {\bf 97},
  5607--5615 (1992).

\bibitem{feynm65.03}
R.~P. Feynman, {\em The Feynman Lectures on Phyiscs Volume III\/}
  (Addison-Wesley, 1965).

\bibitem{tripa04.01}
K.~Tripathy, J.~P\'{e}rez-Moreno, M.~G. Kuzyk, B.~J. Coe, K.~Clays, and A.~M. Kelley,
  \enquote{{Why hyperpolarizabilities Fall Short of the Fundamental Quantum
  Limits},} J. Chem. Phys. {\bf 121}, 7932--7945 (2004).

\bibitem{clays91.01}
K.~Clays and A.~Persoons, \enquote{{Hyper-Rayleigh Scattering in Solution},}
  Phys. Rev. Lett. {\bf 66}, 2980--2983 (1991).

\bibitem{clays92.01}
K.~Clays and A.~Persoons, \enquote{{Hyper-Rayleigh Scattering in Solution},}
  Rev. Sci. Instrum. {\bf 63}, 3285--3289 (1992).

\bibitem{kieli71.01}
S.~Kielich and R.~Zawodny, \enquote{Tensor Relationship of the Molecular
  Electrc Multipole Moments for all Point Group Symmetries,} Chem Phys. Lett.
  {\bf 12}, 20--23 (1971).

\bibitem{Olbre98.01}
G.~Olbrechts, R.~Strobbe, K.~Clays, and A.~Persoons, \enquote{High-frequency
  demodulation of multi-photon fluorescence in hyper-Rayleigh scattering,} Rev.
  Sci. Instrum. {\bf 69}, 2233 (1998).

\bibitem{xu96.02}
C.~Xu and W.~W. Webb, \enquote{{Measurement of two-photon excitation cross
  sections of molecular fluorophores with data from 690 to 1050 nm},} J. Opt.
  Soc. Am. {\bf 13}, 481--491 (1996).

\bibitem{Oulia01.01}
D.~A. Oulianov, I.~V. Tomov, A.~S. Dvornikov, and P.~M. Rentzepis,
  \enquote{Observations onto the measurement of two-photon absorption
  cross-sections,} Opt. Commun. {\bf 191}, 235--243 (2001).

\bibitem{suthe05.01}
R.~L. Sutherland, M.~C. Brant, J.~Heinrichs, J.~E. Rogers, J.~E. Slagle, D.~G.
  McKean, and P.~A. Fleitz, \enquote{Excited-state characterization and
  effective three-photon absorption model of two-photon-induced excited-state
  absorption in organic push–pull charge-transfer chromophores,} J. Opt. Soc.
  Am. B {\bf 22}, 1939--1948 (2005).

\bibitem{belfi07.01}
K.~D. Belfield, M.~Bondar, F.~E. Hernandezt, O.~V. Przhonska, and Y.~S.,
  \enquote{Two-photon absorption cross section determination for fluorene
  derivatives: Analysis of the methodology and elucidation of the origin of the
  absorption processes,} J. Phys. Chem. B {\bf 111}, 12\,723--12\,729 (2007).

\bibitem{kuzyk98.01}
M.~G. Kuzyk and C.~W. Dirk, {\em Characterization techniques and tabulations
  for organic nonlinear optical materials\/} (Marcel Dekker, 1998).

\bibitem{fu07.01}
J.~Fu, A.~P. Lazaro, D.~J. Hagan, E.~W. Van~Stryland, O.~V. Przhonska, M.~V.
  Bondar, Y.~L. Slominsky, and A.~D. Kachkovski, \enquote{Molecular structure
  -- two-photon absorption property relations in polymethine dyes,} J. Opt.
  Soc. Am. B {\bf 24}, 56--66 (2007).

\bibitem{fu07.02}
J.~Fu, A.~P. Lazaro, D.~J. Hagan, E.~W. Van~Stryland, O.~V. Przhonska, M.~V.
  Bondar, Y.~L. Slominsky, and A.~D. Kachkovski, \enquote{Experimental and
  theoretical approaches to understanding two-photon absorption spectra in
  polymethine and squaraine molecules,} J. Opt. Soc. Am. B {\bf 24}, 67--76
  (2007).

\end{thebibliography}

\clearpage

\end{document}